\documentclass[aps,twocolumn,showpacs,prl,superscriptaddress,unsortedaddress]{revtex4}
\usepackage{epsf}
\usepackage{graphicx}
\usepackage{tabularx, booktabs, makecell, amsmath}
\newcommand{\etal}{{\it et al.}}

\begin{document}

\title{Spectroscopic evidence for the convergence of lower and upper valence bands of PbQ (Q=Te, Se, S) with rising temperature}
         
\author{J. Zhao}
\affiliation{Department of Physics, University of Virginia, Charlottesville, VA 22904}
\author{C. D. Malliakas}
\affiliation{Department of Chemistry, Northwestern University, Evanston, IL 60208}
\affiliation{Materials Science Division, Argonne National Laboratory, Argonne, IL 60439}
\author{N. Appathurai}
\affiliation{Department of Physics, University of Virginia, Charlottesville, VA 22904}
\author{V. Karlapati}
\affiliation{Department of Physics, University of Virginia, Charlottesville, VA 22904}
\author{D. Y. Chung}
\affiliation{Materials Science Division, Argonne National Laboratory, Argonne, IL 60439}
\author{S. Rosenkranz}
\affiliation{Materials Science Division, Argonne National Laboratory, Argonne, IL 60439}
\author{M. G. Kanatzidis}
\affiliation{Department of Chemistry, Northwestern University, Evanston, IL 60208}
\affiliation{Materials Science Division, Argonne National Laboratory, Argonne, IL 60439}
\author{U. Chatterjee}
\affiliation{Department of Physics, University of Virginia, Charlottesville, VA 22904}

\date{\today}

\begin{abstract}
We have conducted  temperature dependent  Angle Resolved Photoemission Spectroscopy 
(ARPES) studies of the electronic structure of  PbTe, PbSe and PbS. Our ARPES measurements  provide direct evidence for the  \emph{light}  hole upper valence bands (UVBs) and the so-called \emph{heavy} hole lower valence bands (LVBs), and an unusual temperature dependent relative movement between their band maxima leading to a monotonic decrease in the energy separation  between LVBs and UVBs with increase in temperature. This enables convergence of these valence bands and  consequently an effective increase in the valley degeneracy in PbQ at higher temperatures, which has long been believed  to be the driving factor behind their extraordinary thermoelectric performance.
\end{abstract}

\maketitle
 The unique electronic structures  of lead chalcogenides PbQ (Q=Te, Se, S) have made them canonical systems for fundamental studies of thermoelectric (TE) properties \cite{Vineis2010, Snyder2012, Dresselhaus2007}. Recently, new concepts of ``all scale hierarchical architecture processing" \cite{Kanatzidis2014,Kanatzidis2010, Snyder2011, Johnsen2011} have lead to significant advancements in their TE performance. For instance, $p$-type nanostructured PbTe holds the current performance record for high temperature energy conversion \cite{Girard2011, Pei2011, Biswas2011a, Biswas2011b}. Although these materials have been studied for  decades, they consistently surprise us with new physical phenomena. One such example is the recently reported unexpected appearance of local Pb off-centering dipoles  on warming without a structural transition  in PbTe  \cite{Bozin2010}, which results in an unusual anharmonicity of the bonds and subsequently, giant phonon-phonon scattering at high temperatures. Moreover, these systems have recently been shown to host  various novel quantum states of matter  such as topological crystalline insulator \cite{Liang Fu, Hasan_PbTe, Ando_PbTe, Dziawa_PbSe, Vidya_PbSe} in Pb$_{1-x}$Sn$_x$Se and Pb$_{1-x}$Sn$_{x}$Te, and superconductivity with close connection  to charge Kondo anomaly in the normal state  of Tl doped PbTe \cite{Ian Fisher}. 


The valence band structure of PbQ is intricate--there are hole bands, known as upper valence bands (UVBs),  with maxima at  $L$ points, and extended electronic states, presumably due to  flat secondary valence bands, referred  to as lower valence bands (LVBs). These secondary electronic states occur along $\Gamma$- K  and  $\Gamma$-X lines with maxima at lower energies  compared to the UVBs \cite{UVB_LVB_1, UVB_LVB_2, UVB_LVB_3, UVB_LVB_4}. Despite elaborate investigations on PbQ, the mechanism behind their high TE efficiency and in particular, the connection between temperature ($T$) evolution of their thermopower and  electronic structure is  controversial. 

There are a number of reports in which $T$ dependent thermopower of these materials has been interpreted in terms of a two-band analysis involving  UVBs and LVBs  \cite {Heremans2013, Snyder2013, Pei2011, Biswas2011b, Kolomoet1968, UVB_LVB_2,Allgaier1968} and a $T$ dependent relative shift between these bands leading to their eventual crossover at certain characteristic $T$'s. However, there is no direct experimental evidence for such a change in valence band structure with $T$. The high $T$ enhancement  of thermopower in PbTe and PbSe is attributed  to the dominant contribution from  highly degenerate LVBs, associated with an effective mass heavier than that of the UVBs.  Even though the salient features of the findings in these studies are similar, the reported values of the crossover $T$'s are markedly different  from each other. For example, early reports going back to the 60s and also some of the later ones concluded a crossover of two valence bands in PbTe at $T\sim$450K. This view prevailed in the literature until it was shown a year ago to happen at a much higher $T$ $\sim$750K by the results of magnetic field dependent Hall coefficient measurements at elevated temperatures \cite {Heremans2013}. On the  contrary, recent reports based on first principal calculations cast serious doubts on  the very notion of such  two-band interpretation of thermopower data \cite{Ekuma2012, DJS} and in particular, on the models based on $T$ dependent electronic structure. Taking all these into consideration, an in-depth examination of momentum dependent electronic structure of these compounds as a function of $T$ using ARPES \cite{HUFNER}, which has the unique capability to simultaneously probe energy and momentum of the occupied electronic states in a solid, is highly desirable. 

A number of ARPES experiments \cite{OLD_ARPES, NAKAYAMA, Hasan_PbTe, Ando_PbTe, Dziawa_PbSe}  have indeed been performed on PbTe and PbSe, but the emphasis  has been on completely different  aspects of their electronic structures. 
In this article, we focus on $T$ dependent  ARPES investigations of  valence bands in PbQ and resolve the following lingering issues: What is the valence band structure of these materials? How does rising temperature impact the valence band structure? Our main findings are as follows: (i) there are indeed two distinct valence bands, i.e., LVBs and UVBs, in these materials, (ii) the LVBs lie deeper in energy in comparison with the UVBs (iii) the energy separation between their band edges at $T \sim100$K is largest in PbS, smallest in PbTe, while intermediate in PbSe, and (iv) these two bands do change with $T$ such that the energy separation between their band maxima continuously decreases with increasing $T$ in all three compounds .

\begin{figure}
\includegraphics[width=3.5in]{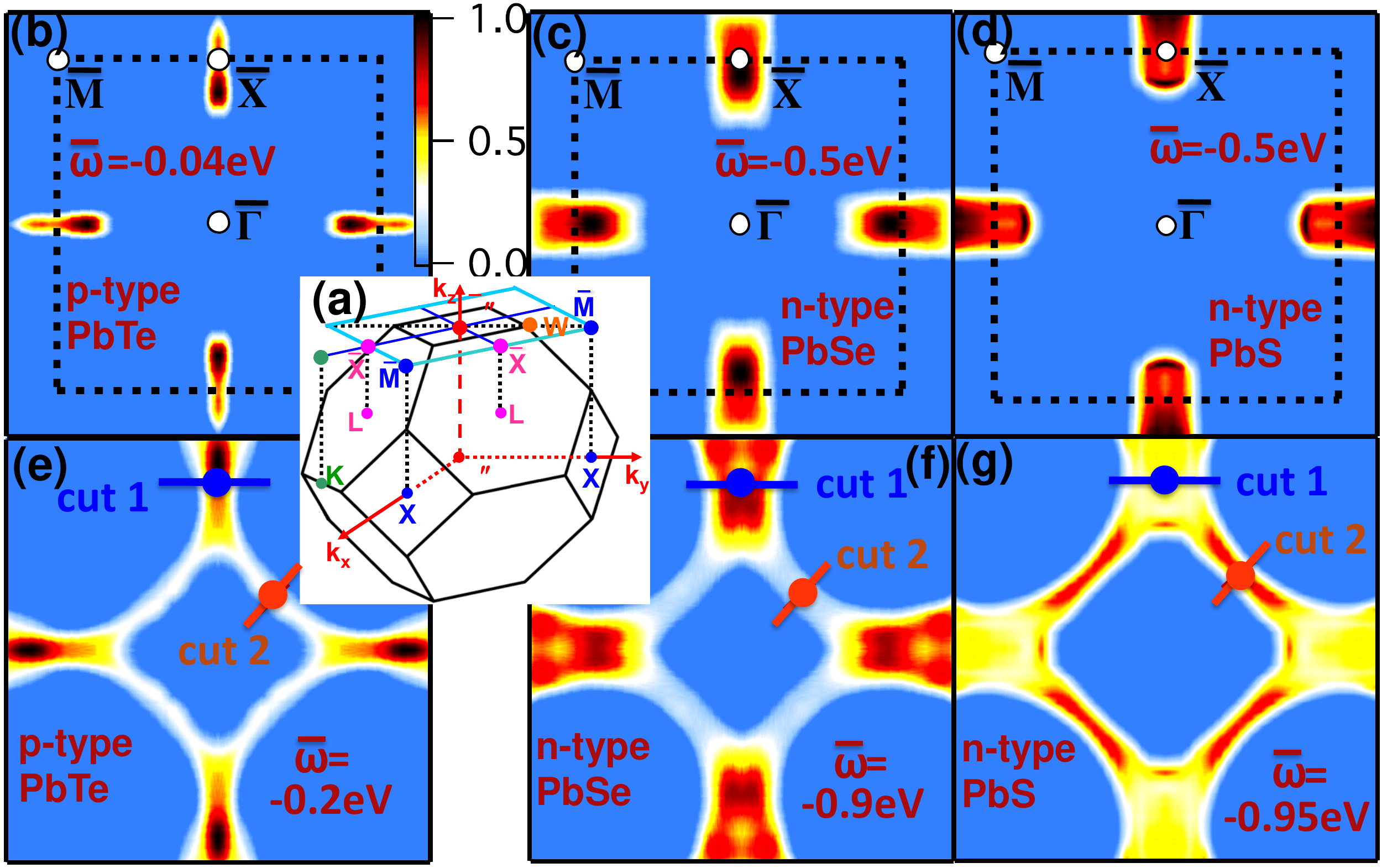}
\caption{ (Color online) (a) Bulk BZ (black lines) associated with a fcc crystal structure and corresponding (001) surface BZ (sky-blue lines).  Constant-energy ARPES intensity maps of a $p$- type PbTe sample at (b) $\overline{\omega}=-40$ meV and (e) $\overline{\omega}=-200$ meV. (c, f) are analogous plots for $n$- type PbSe at $\overline{\omega}=-500$ meV and -900 meV respectively, while (d, g) for $n$- type PbS at $\overline{\omega}=-500$ meV and -950 meV respectively. Blue/ Red lines in (e, f, g) denote Cut1/Cut2 along which ARPES data have been taken to construct Figs. 2a$-$2c /Figs. 2d$-$2f. Blue/Red dots correspond to the momentum locations of the top of the UVBs/LVBs.}
\label{FIG1.pdf}
\end{figure}
We have carried out $T$ dependent ARPES experiments on a number of $n$- and $p$- type PbQ single crystal samples at the PGM beamline of Synchrotron Radiation Center, Stoughton, Wisconsin using a Scienta R4000 electron analyzer. ARPES measurements were performed using 22 eV photon energy and data were collected at 2 and 4 meV energy intervals. The energy and momentum resolutions were approximately  20 meV and 0.0055 $\AA^{-1}$  respectively. PbQ samples were prepared by melting mixtures of Pb and Q at 100$-$150K above the individual melting points of PbQ inside evacuated fused silica tubes. PbI$_2$ was used for achieving  $n$-type doping, while Na for $p$-type. Typical carrier concentrations of the $n$ and $p$-type samples ranged from 2$-$5$\times10^{19} \text{cm}^{-3}$ and 0.2$-$2$\times10^{19} \text{cm}^{-3}$ respectively. These samples were cleaved $\it {in} {situ}$ to expose fresh surface (001) of the crystal for ARPES measurements. They were cleaved  both at low and high temperatures to check consistency in the $T$ dependent data and similar results were found. The chemical potential $\mu$, was obtained by fitting ARPES data from a  polycrystalline gold sample with a resolution broadened Fermi function.

\begin{figure*}
\includegraphics[width=7in]{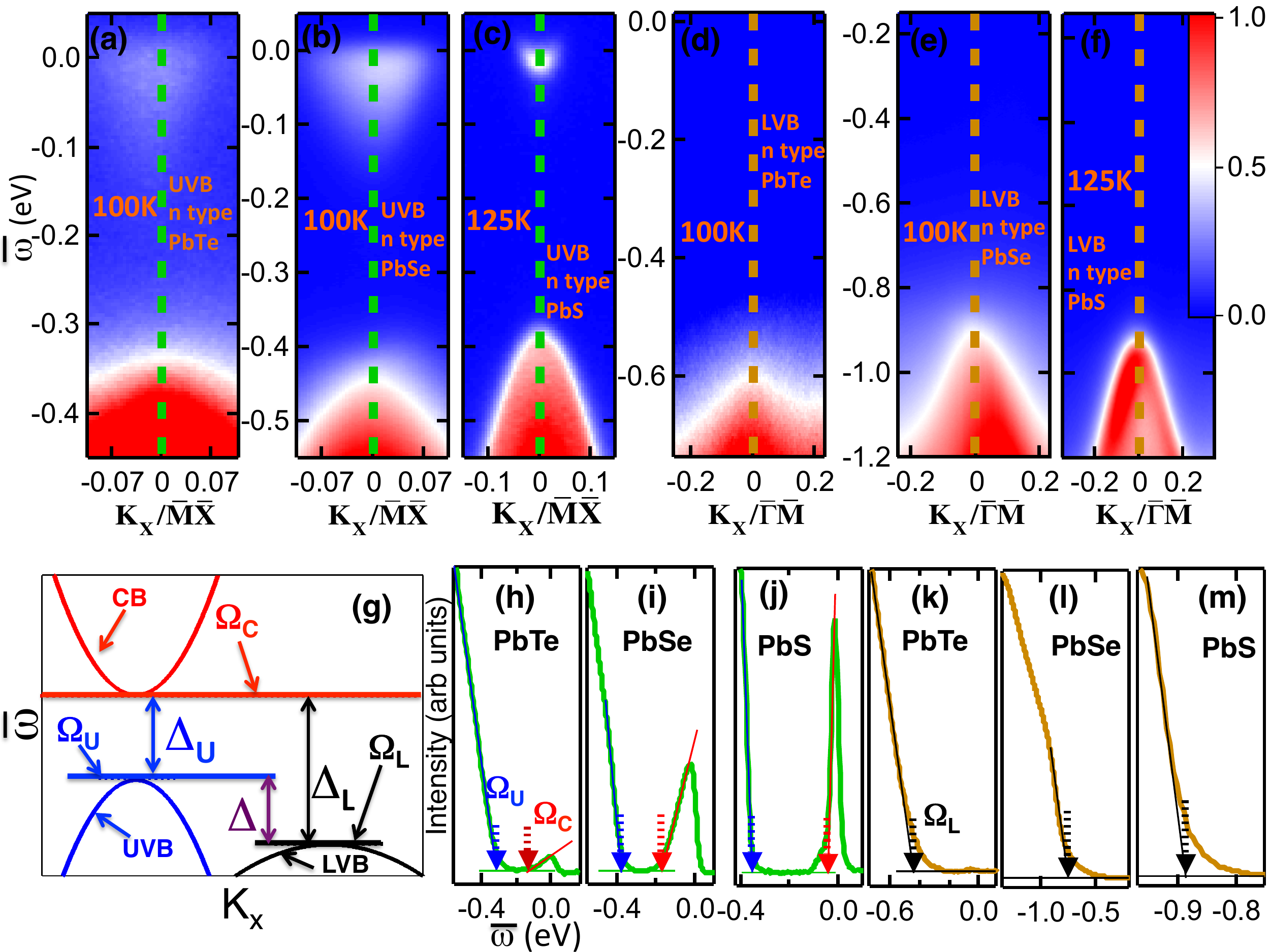}
\caption{ (color online) Dispersion intensity maps along cut 1 for (a)  PbTe, (b) PbSe and (c) PbS, and along cut 2 for (d)  PbTe, (e) PbSe and (f) PbS. (g) A schematic diagram showing relative energy locations of the conduction band (CB), LVB and UVB in PbQ. (h-j) EDCs at the momentum location of the top of the UVB for PbTe, PbSe and PbS respectively.  (k-m) EDCs at the momentum location of the top of the LVB  for PbTe, PbSe and PbS respectively. Blue/Red lines through data points in (h-j) guide almost linear progression of intensity of the EDCs with $\overline{\omega}$ below/above its relatively sharp upsurge at $\overline{\omega}=\Omega_{\text U}$/$\Omega_{\text C}$ indicated by blue/red arrows. Black lines and arrows in (k-m) mark  similar features of the EDCs  associated with LVBs at $\Omega_{\text L}$. (n) A table containing values of $\Delta_{\text L}$, $\Delta_{\text U}$ and $\Delta$ obtained from  $n$-type PbTe, PbSe and PbS samples at $T\sim 100$K.}\label{FIG2.pdf}
\end{figure*}

In order to elucidate  ARPES data, we first consider the bulk Brillouin Zone (BZ) of  PbQ, which has a face centered cubic (fcc) crystalline structure. This BZ is represented in Fig. 1a. Since PbQ is preferentially cleaved along (001) plane, one needs to pay attention to  its square surface BZ projected onto the (001) plane for interpreting  ARPES data. As shown in Fig. 1a, the $\Gamma$ and $L$ points project on the  ($\overline{\Gamma}$) and $(\overline{X})$ of the surface BZ. In Figs. 1b$-$1g, we show constant-energy ARPES intensity maps $I(\bf{k_{x}},  \bf{k_{y}}, \overline{\omega}\text{=constant})$, data as a function of in-plane momentum components $\bf{k_{x}}$ and $\bf{k_{y}}$ at a fixed  $\overline{\omega}$, where $\overline{\omega}$ is electronic energy with respect to $\mu$. Figs. 1b and 1e correspond to $\overline{\omega}=$ -40 meV and  $\overline{\omega}$= -200 meV respectively for a $p$- type PbTe sample.  As expected, $L$-centered hole pockets derived from UVBs are clearly visible around $\overline{X}$  at $\overline{\omega}$= -40 meV (Fig. 1b). When  $\overline{\omega}$ is changed to -200 meV, there is  a remarkable change in topology of  the intensity map via appearance of tube like regions connecting these isolated pockets (Fig. 1e). 
These tubular regions at higher $\overline{\omega}$ have been predicted in several band structure calculations \cite{DJS, SVANE, FS_PBTE} and have also been seen in recent ARPES measurements on Pb$_{1-x}$Sn$_{x}$Te \cite{Hasan_PbTe} and SnTe \cite{LITTLEWOOD}. Qualitatively, similar evolution of intensity maps with $\overline{\omega}$ are observed in Figs. 1c and 1f  for PbSe, and in Figs. 1d and 1g for PbS.

To further investigate the valence band structure of these materials, we concentrate  on Fig. 2. Here electronic dispersions along two specific momentum space cuts, namely cut 1 and cut 2 defined in Fig. 1, are presented. Fig. 2a$-$2c correspond to dispersion  ($\overline{\omega}$ vs $\bf{k_{x}}$) intensity maps along cut 1, while Figs. 2d$-$2f to those along cut 2. One can recognize an electron band, the conduction band, separated in energy and momentum from a hole band, the UVB, in each intensity map along cut 1. On the contrary, a hole band alone is present  along cut 2. By comparing data along cut 1 and cut 2, it can be immediately realized that the hole bands along cut 1, the UVBs,  are distinct from those along cut 2, which we identify as the so-called LVBs. These clearly corroborate the two-band picture of PbQ.

 Now, we will determine  band gaps associated with LVBs and UVBs in various PbQ samples by examining Energy Distribution Curves (EDCs) \cite{HUFNER} at momentum locations where these hole bands reach their individual maxima (Figs. 2h$-$2m). In this context, an EDC is the distribution of electrons as a function of $\overline{\omega}$ at a given $\bf{k_{x}}$. For the purpose of quantifying band gaps corresponding to  a LVB or a UVB in a sample, we need to adopt some method of detecting not only the $\overline{\omega}$ positions of their maxima, but also the $\overline{\omega}$ position of the bottom of the conduction band (Fig. 2g). We illustrate this in Fig. 2h. It is easy to notice that the intensity of the EDC below a certain value of $\overline{\omega}$ builds up  abruptly on top of its background value and from then on it continues to grow almost linearly.  We define this particular $\overline{\omega}$ as $\Omega_{\text{U}}$ and this marks the $\overline{\omega}$ location of the top of the UVB.  Similarly, one can  also note a relatively sharp rise in intensity on top of the  background above a certain value of $\overline{\omega}$. We refer this $\overline{\omega}$ as  $\Omega_{\text C}$ and this corresponds to the bottom of the conduction band. Once we determine $\Omega_{\text C}$ and $\Omega_{\text U}$, it is straightforward to evaluate the band gap $\Delta_{\text U}$ of the UVB using  $\Delta_{\text U}$=$\Omega_{\text C}$-$\Omega_{\text U}$ (Fig. 2g). Following the same procedure we locate $\Omega_{\text L}$, the $\overline{\omega}$ position for the top of a LVB, and calculate the relevant band gap $\Delta_{\text L}$ using $\Delta_{\text L}$=$\Omega_{\text C}$-$\Omega_{\text L}$ (Fig. 2g). Since each of these samples does not have the exact same carrier concentration, the useful quantity to be compared is $\Delta$=$\Delta_{\text L}$-$\Delta_{\text U}$, or in other words the energy separation between the maximum of a LVB and that of a UVB. Table I  demonstrates that $\Delta$ is largest for PbS ($\sim 536$ meV) and smallest for PbTe ($\sim130$ meV), while in between ($\sim 394$ meV) for PbSe at $T\sim100$K. 
\begin{figure}
\includegraphics[width=3.5in]{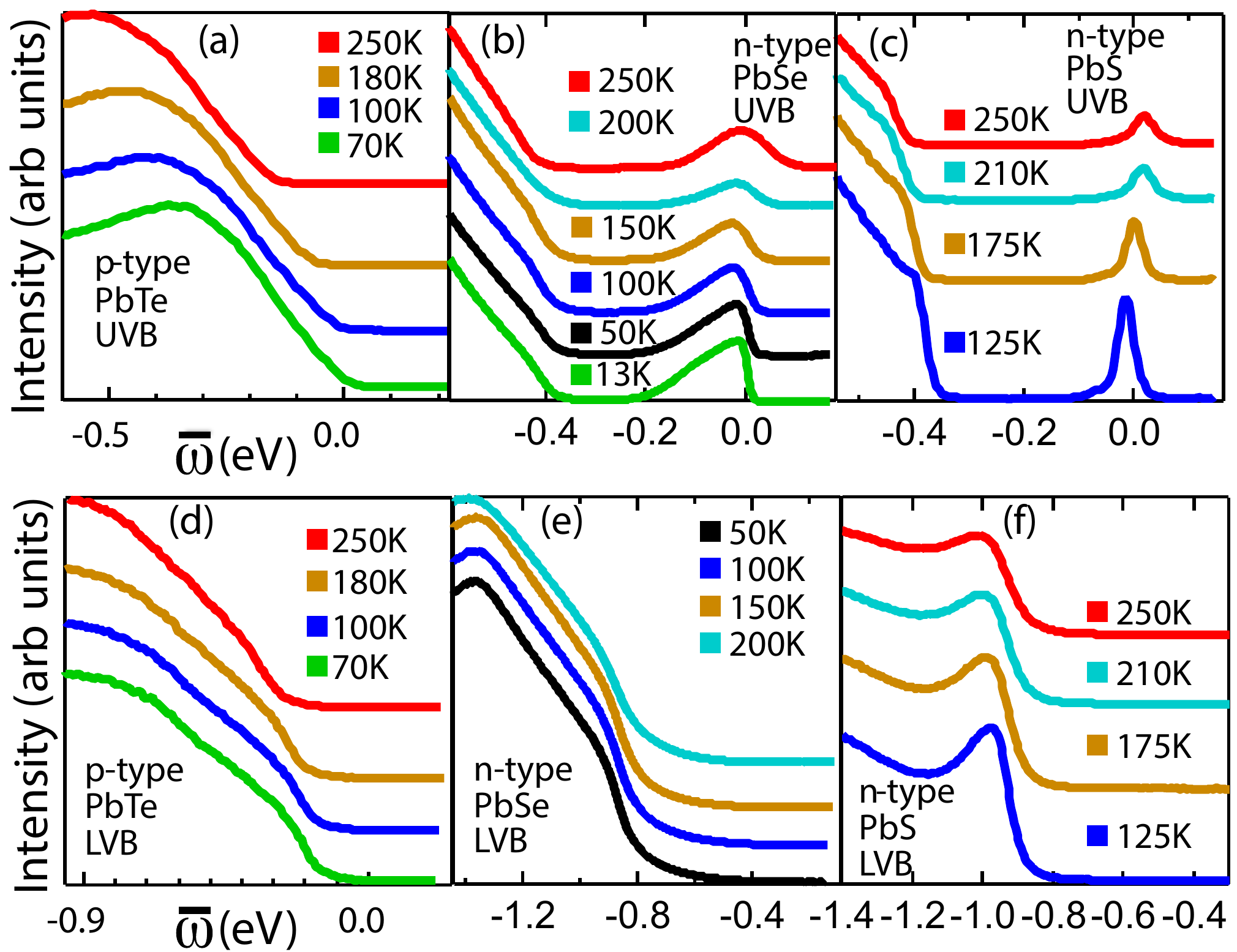}
\caption{ (color online) EDCs as a function of $T$ at the momentum location associated with the top of the UVB for (a) $p$- type PbTe, (b) $n$-  type PbSe and (c) $n$-  type PbS, while those at the top of the LVB for for (d) $p$- type PbTe, (e) $n$-type PbSe and (f) $n$- type PbS.}
\label{FIG3.pdf}
\end{figure}

\begin{table}
\caption{Values of $\Delta_{\textbf{L}}$, $\Delta_{\textbf{U}}$, and $\Delta$ obtained from $n$-type PbTe, PbSe, and PbS samples at $T \sim$ 100K}
\centering
\renewcommand{\arraystretch}{1.5}
\begin{tabularx}{.48\textwidth}{>{\centering\arraybackslash}X >{\centering\arraybackslash}X >{\centering\arraybackslash}X >{\centering\arraybackslash}X}\hline\hline
Materials & {$\Delta_{\textbf{U}}$ (meV) }& {$\Delta_{\textbf{L}}$ (meV) }& {$\Delta$ (meV)}\\
\hline
PbTe & 190 & 320 & 130\\   
PbSe & 206 & 600 & 394\\   
PbS & 310 & 846 & 536\\    \hline\hline
\end{tabularx}
\end{table}

The objective of the remainder of the paper is to interrogate  the impact of increasing  $T$ on LVBs and UVBs. This  is summarized in Figs. 3 and 4. In Figs. 3b and 3c we plot EDCs  for $n$- type PbSe and PbS samples respectively as a function of $T$ at the  momentum locations of the top of their individual UVBs, while in  Figs. 3e and 3f at those of their LVBs. We also display similar plots for the $p$- type PbTe sample in Figs. 3a, 3d.  Following the procedures  described in Fig. 2, we obtain $\Delta_{\text{U}}$ and $\Delta_{\text{L}}$ as a function of $T$ for both PbSe and PbS and then plot them in Figs. 4a, 4b. Although  $\Delta_{\text{L}}$ does not change appreciably with $T$, $\Delta_{\text{U}}$ grows almost linearly with rising $T$, which is consistent with positive temperature coefficients of band gap  found by optical experiments  in PbQ \cite{Miller1961, Tauber1966, Tsang1971, Snyder2013}. It is worth mentioning that such positive rate of change of band gap with $T$ in PbQ is contrary to what occurs in a majority of semiconductors. In fact, this \textit{``anomaly"} helps PbQ to achieve high thermoelectric efficiency since it can mitigate the bipolar effects from intrinsic carrier activation. The later is responsible for the suppression of thermoelectric figure of merit $zT$ at high $T$ in a material. We  further point out that $\Omega_{\text C}$ for the $p$- type PbTe sample  can't  be  determined since its conduction band lies in the un-occupied side of the band structure. We can, however, obtain 
$\Omega_{\text U}$ and $\Omega_{\text L}$ from Fig. 3a and Fig. 3d respectively and thus, plot $\Delta$=$\Omega_{\text U}$-$\Omega_{\text L}$ as a function of $T$ for PbTe along with PbSe and PbS in Fig. 4c.
                                                                                                                                                                                                                                                                                                                                                                                                                                                                      \begin{figure}
\includegraphics[width=3.4in]{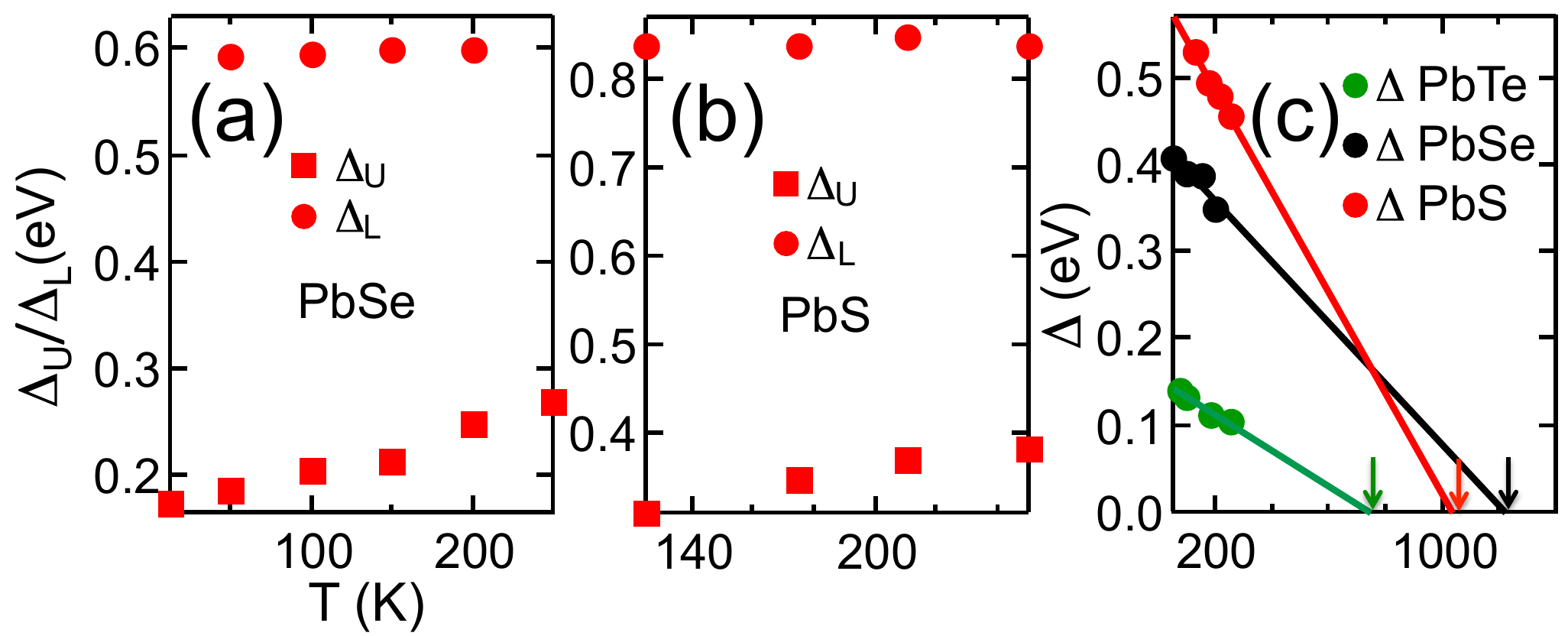}
\caption{ (color online) $T$ dependence of $\Delta_{\text{U}}$ and $\Delta_{\text{L}}$ for (a) PbSe and (b) PbS. For both PbSe and PbS, $\Delta_{\text{U}}$ grows  with $T$, while $\Delta_{\text{L}}$ remains almost constant. (c) $\Delta$ as a function of $T$ for PbSe, PbS and PbTe. Green, black  and red lines are linear fits to the data points for PbTe, PbSe and PbS respectively. Green, black and red arrows indicate corresponding $T^*$ values.}   
\label{FIG4.pdf}
\end{figure}

It is easy to infer that $\Delta$ for each PbQ sample decreases monotonically with increasing $T$ (Fig. 4c). Data points within the measured $T$ values  can be well represented by  straight lines. If we assume that this linear change of $\Delta$ with $T$ for each sample persists even above the highest measured temperature, we can extrapolate these straight lines and conclude that at certain characteristic temperature $T^*$, $\Delta$ will reduce to zero. This means that  the band edge of the UVB will cross through that of the LVB at $T^*$. An examination of Fig. 4c implies that $T^*\sim 750$K, $1200$K and $1000$K for PbTe, PbSe and PbS respectively.  Although our estimation of $T^*$ involves an extrapolation over a large $T$ range, surprisingly the $T^*$ value for PbTe obtained from our ARPES experiments matches reasonably well with $T^*$ value from recent magnetic and optics measurements \cite {Heremans2013, Snyder2013}. Moreover, the $T^*$ value obtained for PbS also agrees  well with that in Ref. 24, however, there is  slight discrepency in case of PbSe. All these results suggest that PbTe, PbSe and PbS should become  semiconductors with indirect band gap for $T>T^*$, where the heavy hole LVB rises in energy above the light hole UVB. In this scenario, the charge transport in these materials should be dominated by the heavy holes created due to thermal excitations as the temperature approaches $T^*$ and consequently, the energy gap between LVB and UVB becomes comparable with k$_{\text B}T$.

U.C. acknowledges support from the startup fund provided by the University of Virginia. Work at Argonne National Laboratory (C.D.M., D.Y.C., S.R., M.G.K.) was supported by the U.S. Department of Energy, Office of Basic Energy Sciences, Division of Materials Science and Engineering. U.C. thanks S.D. Mahanti for helpful discussions.

\end{document}